\documentclass[doublecol]{epl2}

\usepackage{ifthen}
\usepackage{ifpdf}
\usepackage{color}

\ifpdf
\usepackage{graphicx}
\usepackage{epstopdf}
\else
\usepackage{graphicx}
\usepackage{epsfig}
\fi

\usepackage{latexsym}
\usepackage{amsmath}
\usepackage{amssymb}
\usepackage{bm}
\usepackage{wasysym}

\newcommand{\abs}[1]{\left|#1\right|}

\newcommand{\df}{{\rm d}}

\newcommand{\const}{\mbox{const}}

\newcommand{\bra}{\left\langle}
\newcommand{\ket}{\right\rangle}

\newcommand{\mass}{\mathsf{m}}

\newcommand{\tbox}[1]{\mbox{\tiny #1}}
 
\newcommand{\mylabel}[1]{\label{#1}} 
\newcommand{\beq}{\begin{eqnarray}}
\newcommand{\eeq}{\end{eqnarray}} 
\newcommand{\be}[1]{\begin{eqnarray}\ifthenelse{#1=-1}
{\nonumber}{\ifthenelse{#1=0}{}{\mylabel{e#1}}}}
\newcommand{\ee}{\end{eqnarray}} 

\newcommand{\Eq}[1]{\textcolor{blue}{Eq.\!\!~(\ref{#1})}} 
\newcommand{\Fig}[1]{\textcolor{blue}{Fig.}\!\!~\ref{#1}} 
\newcommand{\hide}[1]{}
\newcommand{\mycite}[1]{\textcolor{blue}{\cite{#1}}} 

\title{Non-adiabatic pumping in an oscillating-piston model} 
\shorttitle{Non-adiabatic pumping}

\author{Maya Chuchem$^1$, Thomas Dittrich$^{2,3}$, Doron Cohen$^1$}

\institute{
\mbox{$^1$Department of Physics, Ben-Gurion University of the Negev, Beer-Sheva 84105, Israel} \\
\mbox{$^2$Departamento de F{\'\i}sica, Universidad Nacional de Colombia, Bogot\'a D.C., Colombia} \\
\mbox{$^3$CeiBA -- Complejidad, Bogot\'a D.C., Colombia}
}

\abstract{
We consider the prototypical ``piston pump" operating on a ring, 
where a circulating current is induced by means of an AC driving. 
This can be regarded as a generalized {Fermi-Ulam model}, 
incorporating a finite-height moving wall (piston) and non trivial
topology (ring). The amount of particles transported per
cycle is determined by a layered structure of phase-space. 
Each layer is characterized  by a different drift velocity. 
We discuss the differences compared with the adiabatic and Boltzmann pictures, 
and highlight the significance of the "diabatic" contribution that 
might lead to a counter-stirring effect. 
}

\begin{document} 
\maketitle


{\em Stirring} is the operation of inducing a DC circulating current 
by means of AC driving. This is naturally achieved by integrating  
a {\em pump} \mycite{bpt,bpt2,avron} in a closed circuit \mycite{pmc,pMB}.
It can also be regarded as a variation   
of a Hamiltonian {\em ratchet} \mycite{ratchet1,ratchet2}  
where transport is induced in a periodic array.
Pumping and stirring have largely been considered in the regime of
slow (adiabatic) driving, where it can be related
to the Berry phase that is associated with the driving cycle. 
This adiabatic approach is based on a simple picture 
of probability flow.

Challenging this oversimplified view, we argue that there are 
typical circumstances where the analysis should go beyond the adiabatic picture, 
even for very slow driving. 
We here present a detailed account of {\em deterministic} stirring
that naturally extends into the non-adiabatic regime, 
complementary to related studies of  
stochastic stirring \mycite{saar}, 
Brownian ratchets~\mycite{bratchet}, Brownian motors~\mycite{bmotor}, 
stochastic~\mycite{sp} and chaotic~\mycite{cp} pumps. 

We shall show that for a prototype system, the oscillating-piston model, 
even if the driving is very slow, the dynamics is actually complex,  
due to a non-trivial structure of phase-space, 
leading to drastic consequences for the transport.

\section{Outline} 
After introducing the model, we describe the 
expectations that are based on a {\em stochastic Boltzmann picture}, 
and on a {\em deterministic adiabatic picture}.
These suggest two different {\em parametric} results 
for the amount~$Q$ of pumped particles.  
Then we present a proper analysis 
of the mixed phase-space dynamics, 
and highlight the limitations of the traditional reasoning.

\begin{figure}

\includegraphics[clip,width=0.9\hsize]{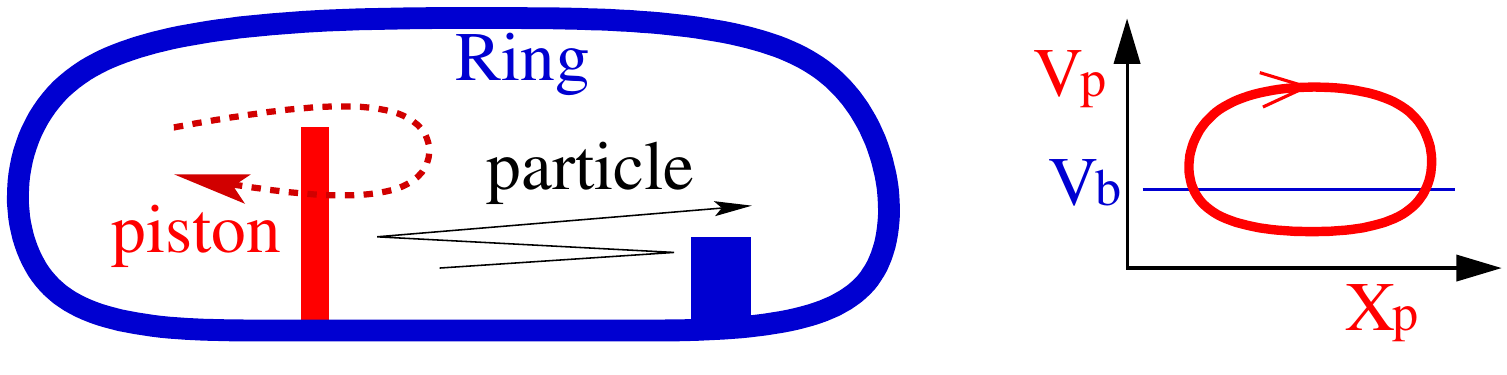}

\caption{
(Color on-line)
An oscillating piston pump is integrated into a ring.
The pumping cycle is illustrated on the right.
The scattering within the ring is modeled as a potential barrier. 
We are considering a closed geometry that can be regarded 
a generalization of the {Fermi-Ulam model}.
If we considered an open geometry, the pumping 
device would be connected between two unbiased 
reservoirs (not presented). }

\label{f1}
\end{figure}

\section{The model}
We consider the prototype system illustrated in \Fig{f1}:
A particle with mass $\mass$ moves in a ring of length~$L$. There is a
fixed barrier~$V_b$ at~$x{=}0$ and a moving wall (piston) 
of height $V_p$ at~$x{=}X_p$. 
In the oscillating piston paradigm \mycite{avron,pmt}, 
the two control parameters of the piston are cycled periodically 
through a closed loop in parameter space.  
The {\em pumping cycle} $(X_p(\varphi),V_p(\varphi))$,
with $\varphi = \Omega t$,  
consists of translating the piston some distance to the right, 
shrinking its ``height", pulling it back to the left, 
and restoring its original ``height". 
In the sequel, we shall assume a harmonic driving with
phase shift $\pi/2$ between the two parameters,
\be{100}
X_p(\varphi) \ &=& \ X_0 - \delta X_p\cos(\varphi), \\ \label{e101}
V_p(\varphi) \ &=& \ V_p + \delta V_p \sin(\varphi).
\eeq
The velocity $v$ of the particle  
in dimensionless units is defined as ${u=v/(\delta X_p\Omega)}$. 
With the same convention the piston velocity~$ \dot{X_p}$
is expressed as ${u_x(\varphi)=\sin(\varphi)}$.
The transmissions of the barrier and the piston 
are given by boolean expressions (true=1, false=0): 
\be{11}
g_b(u) \ &=& \  \Big[|u| > u_b \Big] , 
\\ \label{e12}
g_p(u, \varphi) \ &=& \ \Big[|u{-}u_x(\varphi)| > u_p(\varphi)\Big] ,
\eeq
where $u_{b,p}=[2V_{b,p}/\mass]^{1/2}/(\delta X_p\Omega)$.
In addition to the three dimensionless parameters
${(u_b,u_p,\delta u_p)}$ that describe the barrier and the piston, 
we specify the geometry of the system defining $2\pi\ell_{+} \equiv
X_0/\delta X_p$  and $2\pi\ell_{-} \equiv (L{-}X_0)/\delta X_p$.
A~Poincar\'e section of the dynamics is obtained 
by taking snapshots of $(\varphi,u)$ after each collision 
with the fixed barrier:
\be{5}
\varphi_n' &=&  \varphi_n + \frac{2\pi\ell_n}{|u_n|} , 
\\
u_n' &=&  -u_n + 2u_x(\varphi_n'),  \ \ \  \mbox{if} \ g_p(u_n,\varphi_n'){=}0 ,
\\
\varphi_{n+1}  &=& \varphi_n' + \frac{2\pi\ell_n'}{|u_n'|} ,
\\
\label{e8}
u_{n+1} &=&  -u_n',  \ \ \ \mbox{if} \ g_b(u_n'){=}0 .
\ee
Above $\ell_n$ ($\ell_n'$) is the scaled travel distance  
from the barrier to the piston (from the piston to the barrier). 
In the ``static-wall approximation" it is $\ell_{+}$ or
$\ell_{-}$ depending on the sign of~$u_n$, while in the
simulations the exact value can be numerically determined. 

\section{Objective}
The map above generalizes the {Fermi-Ulam model (FUM)} \mycite{LL}:  
here we have finite heights of piston and barrier, 
and periodic boundary conditions. 
Furthermore, while the FUM has been conceived to study energy absorption 
due to deterministic diffusion in momentum, 
here our interest is in the directed transport 
along the spatial coordinate. 
The amount of particles that are pumped per cycle is:
\be{1}
Q
\ = \ 
\oint I \df t
\ = \ 
\rho_N
\left[
\oint v(\varphi) \frac{\df\varphi}{2\pi}
\right]
\frac{2\pi}{\Omega},
\eeq
where the current ${I \equiv \rho_{N} v(\varphi)}$
at a given moment of time is expressed by the spatial 
density of the particles~$\rho_{N}$, 
and the drift velocity~$v(\varphi)$.
Assuming that a first-order description
of adiabatic transport applies,  
if the piston is displaced with velocity $\dot{X}_p$, 
one expects a well defined induced current ${I = \mathcal{R} \rho_{N} \dot{X}_p}$, 
where $\mathcal{R}$ is an $\Omega$-independent 
dimensionless coefficient that primarily depends on~$V_p$. 
One expects~$\mathcal{R} \to 1$ for ${V_p \to \infty}$  
and ${0<\mathcal{R}<1}$ for finite $V_p$. 
In such a case the amount of particles 
that are pumped per cycle becomes a parametric integral: 
\be{10}
v(\varphi) = \mathcal{R}(\varphi) \ \dot{X}_p
\ \ \ \leadsto  \ \ \  
Q = \rho_N \oint \mathcal{R}(\varphi) \df X_p .
\eeq
Below we try to formulate an adequate picture
of the dynamics that interpolates between the
opposite extremes of exclusively chaotic
and purely regular motion. In particular:
{\bf (i)}  we clarify what is the drift velocity 
for a general non-equilibrium steady state;
{\bf (ii)}  we discuss whether a first-order description
of adiabatic transport applies.
We first review the common expectations.

\begin{figure*}[t]

\hspace*{2cm} (a) \hspace*{8cm} (b) \\
\includegraphics[clip,width=0.425\hsize]{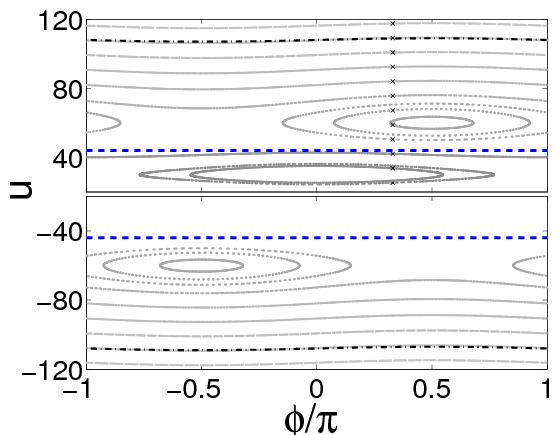}
\hspace*{0.45cm}
\includegraphics[clip,width=0.425\hsize]{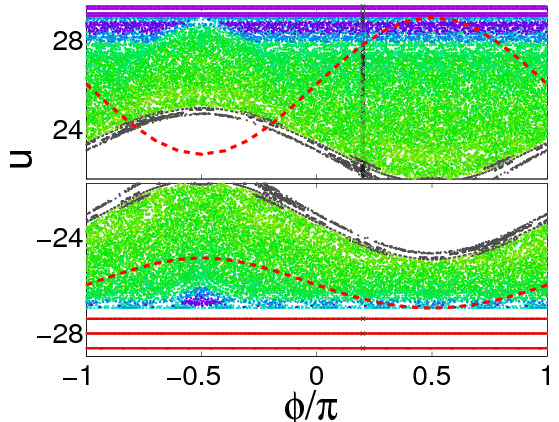} 
\hspace*{0.1cm}
\includegraphics[clip,width=0.1\hsize]{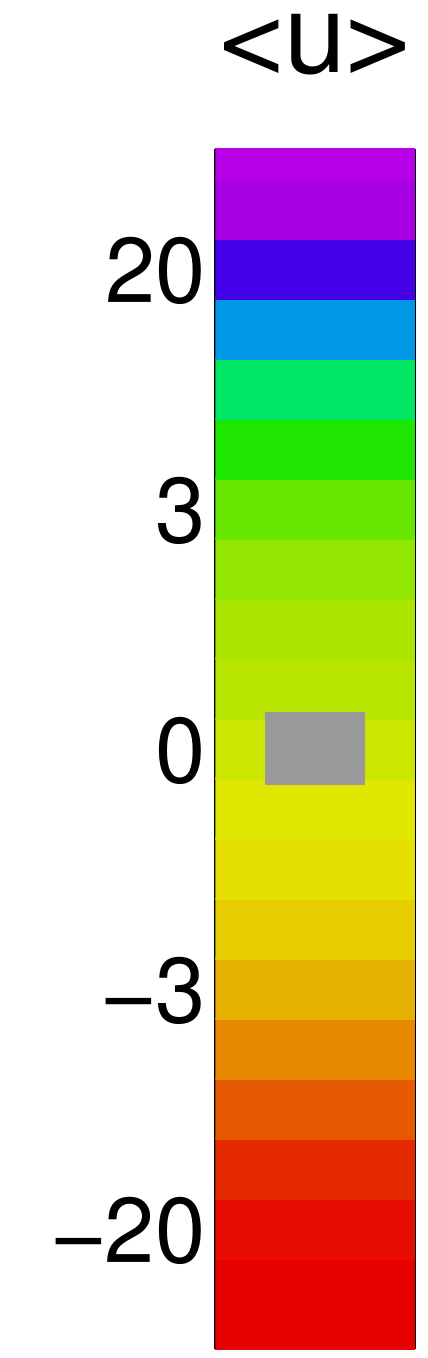} 

\hspace*{2cm} (c) \hspace*{8cm} (d) \\
\includegraphics[clip,width=0.425\hsize]{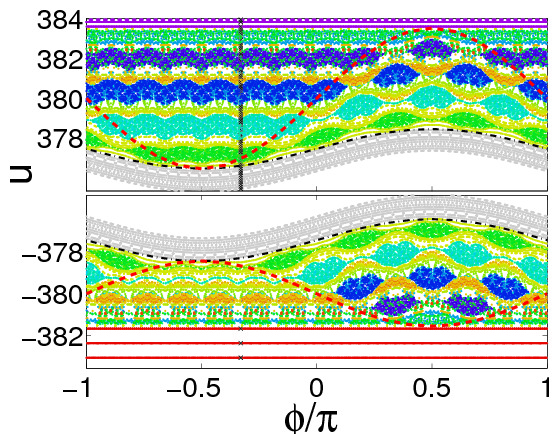}
\hspace*{0.45cm} 
\includegraphics[clip,width=0.425\hsize]{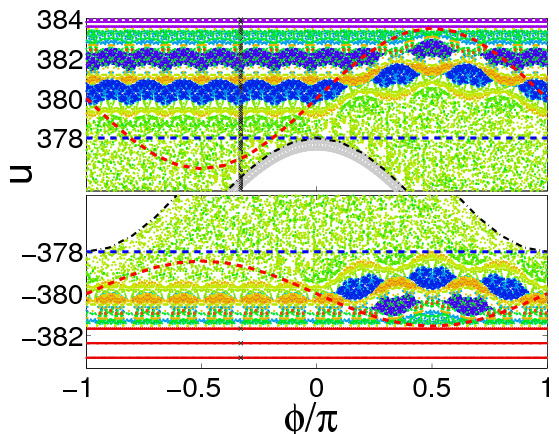}
\hspace*{0.1cm}
\includegraphics[clip,width=0.1\hsize]{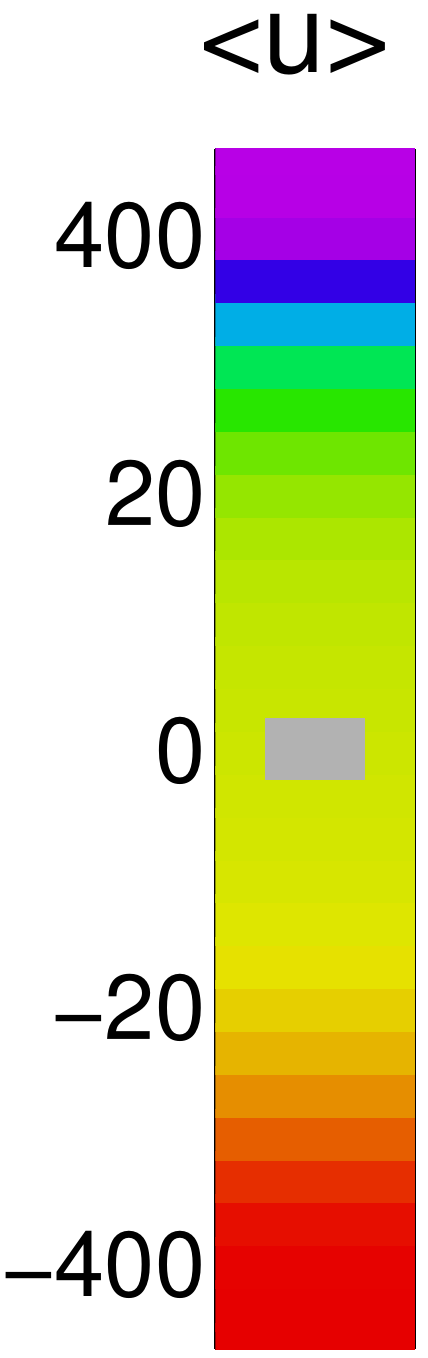}

\caption{The $(\varphi,u)$ Poincar\'e section  
for a particle in a ring, with $l_{\pm}{=}15$, 
considering several cases:
(a)~ring with ${u_p{=}\infty}$ and ${u_b{=}44}$; 
(b)~here ${u_b{=}0}$, and ${u_p{=}26}$, and ${\delta u_p{=}2.04}$;  
(c)~here ${u_b{=}0}$, and ${u_p=380}$ and ${\delta u_p{=}2.55}$;   
(d)~the same but with ${u_b{=}378}$.
The color of each trajectory reflects 
its average drift velocity $\langle u\rangle$.
Non transporting trajectories are colored gray.
The dashed red lines indicate the piston 
scattering threshold~$u^{\pm}(\varphi)$, 
while the dashed blue line gives that of the barrier~$u_b$. 
The dot-dashed black lines are calculated adiabatic trajectories. 
Above the barrier the variation in~$|u|$, due to the bounces with the piston,  
is~$\sin(\varphi)$, with alternating sign in each bounce.
Below the barrier $u$ keeps its sign, and the adiabatic variation 
is~$[\langle |u| \rangle /(2\pi l_{+})]\cos(\varphi)$. 
See the appendix for details regarding the adiabatic trajectories.  
}

\label{f2}
\end{figure*}

\section{The stochastic Boltzmann picture}
The piston stirring problem has been analyzed in the past \mycite{pmt} 
using a stochastic approach with transmissions ${0 < g_b,g_p < 1}$. 
Motivated by the prevailing literature that focuses on 
electronic systems one assumes a Fermi-like energy distribution, 
%
%
such that the initial occupation is 
\be{9}
f(x,p) &=& \frac{1}{2\pi\hbar}, \ \ \ \ \ \ \mbox{for $|p|<\mass
v_{\tbox{F}}$} 
\\ 
\label{e91}
\rho_N &=& \frac{\mass v_{\tbox{F}}}{\pi\hbar}, \ \ \ \ \ \
\mbox{[density of particles]} 
\eeq
where $\mass v_{\tbox{F}}$ is the Fermi momentum.
In a classical context, $\hbar$ can be regarded 
as a parameter that determines the occupation density.  
Of interest is $f(p)=f(0,p)$, the momentum distribution 
at a section ${x=0}$ through which the current is measured. 
Its integral over $\df p$ gives the density of particles~$\rho_N$. 
If ${|\dot{X}_p| \ll v_{\tbox{F}}}$, 
the momentum exchange due to collisions with the piston
affects only a narrow shell of width $\sim 2\mass \dot{X}_p$ around
the Fermi energy ${E_{\tbox{F}}}$. Accordingly one writes 
\beq
I \ = \ \int_{0}^{\infty} \df p  [f(p)-f(-p)] v(p)
 \ = \ 
\rho_N \ \mathcal{R}\dot{X}_p .
\eeq
From here \Eq{e10} is implied, while $\mathcal{R}$ 
is related to the reflectivity of the piston
as follows~\mycite{pmt}:
\be{2}
\mathcal{R}(\varphi) \ \ = \ \ \frac{(1-g_p) \ g_b}{g_p+g_b-2g_pg_b} .
\eeq
In the absence of a barrier (${g_b=1}$) the result is ${\mathcal{R}=1}$. 
If the ring is like a reservoir (${g_b=1/2}$) 
one observes that $\mathcal{R}=1{-}g_p$ is the reflectivity of the piston. 
The latter result is very well known \mycite{avron}, 
conventionally derived using the scattering formalism \mycite{bpt,bpt2}.

\section{Deterministic adiabatic picture}
In our model the transmission coefficients of the barrier are
given by Eqs.(\ref{e11}-\ref{e12}). In \Eq{e2} we have to substitute  the
values at the Fermi energy ${E_{\tbox{F}}=(1/2)\mass v_{\tbox{F}}^2}$,
which gives ${\mathcal{R}(\varphi)=\{0 \,\mbox{or}\, 1\}}$  depending
on whether the piston is ``below" or ``above"  the Fermi energy. 
In the latter case, from \Eq{e10} with \Eq{e91}, 
we get ${dQ=[{\mass v_{\tbox{F}}}/{\pi\hbar}]\df X_p}$,  
which coincides with \mycite{avron}.
If the Fermi energy is ``above" the piston during the whole cycle, 
meaning that ${V_p(\varphi) < E_{\tbox{F}}}$ for any $\varphi$, 
we find ${Q=0}$.  
However, in the latter case, there is a naive (wrong) picture 
that suggests a finite result:
Assuming that~$Q$ is determined by the fraction 
of particles that are affected by the motion of 
the piston, the effective Fermi energy is $V_p(\varphi)$, and hence  
\be{160}
Q \ = \ 
\frac{1}{\pi\hbar} \oint \sqrt{2\mass V_p(\varphi)} \ \df X_p.
\eeq
This ``area" that is enclosed by the cycle 
resembles the {\em action integral} that is encountered 
in the canonical adiabatic picture.  
It has the form of \Eq{e10} with a modified 
`reflection' coefficient ${\mathcal{R}(\varphi)=[V_p(\varphi)/E_{\tbox{F}}]^{1/2}}$.
In spite of the wrong reasoning, \Eq{e160} 
is interesting because it can be justified as an approximation 
to what we call later ``adiabatic contribution".

\section{Non-adiabatic deterministic dynamics}
The Poincar\'e section for the generalized FUM Eqs.(\ref{e5}-\ref{e8})
is illustrated in \Fig{f2}. We indicate there (by dashed red lines) 
the threshold velocity for piston reflection 
%
%
\beq
u_p^{\pm}(\varphi) \ \ = \ \ u_x(\varphi) \pm u_p(\varphi)
\eeq
which is implied by \Eq{e12}.
We define $v^{(\pm)}$ as the velocities that correspond  
to $\max[u_p^{+}]$ and $\min[u_p^{-}]$, respectively, 
and denote by $E^{(\pm)}$ the associated kinetic energies.
For simplicity of presentation we assume that $V_b$ is smaller
than~$E^{(-)}$, which is always smaller than~$E^{(+)}$. Accordingly, 
the ballistic motion of clockwise moving particle   
is not affected if ${E>E^{(-)}}$, while for anticlockwise moving 
particles the condition is ${E>E^{(+)}}$. 

The non-integrable region in phase-space occupies 
the rectangular strip ${[v^{(-)},v^{(+)}]}$. 
One expects adiabatic dynamics if the slowness 
condition $|u|\gg\ell$ is satisfied there. 
Looking at \Fig{f2}cd one observes that this region consists of {\em layers}.
In each layer the motion is chaotic, with the possible exception 
of regular motion in some small islands. 
The lower layers, labeled collectively as ${r=0}$, 
are non-transporting: 
either the particle is bounded in the $\ell_{+}$ 
or in the $\ell_{-}$ segment, or else it occupies the whole ring 
without being ever able to cross one of the barriers.
The motion in the subsequent transporting layers, 
labeled ${r=1,2,...r_c}$, is characterized by a non-zero 
drift velocity 
\beq
\langle v \rangle_r \ \ \equiv \ \ \bar{u}_r \ \Omega \ \delta X_p
\eeq

\section{Stirring}
Define $E^{(0)}$ as the minimal energy required 
for transporting motion. 
Consider a zero temperature Fermi occupation
as the preparation. If either ${E_{\tbox{F}}}$ is below ${E^{(0)}}$  
or above ${E^{(+)}}$ the induced current would be zero. 
In the latter case the non-regular transporting motion is merely 
a ``musical chair" dynamics that takes place deep in the Fermi sea. 
The current would be non-zero if ${E^{(0)}< E_{\tbox{F}} <E^{(+)}}$, 
meaning that only a fraction of the transporting region is occupied.  
The amount of particles that are pumped during 
a period is  ${Q = (2\pi/\Omega)\rho_N \langle v \rangle}$, 
where the cycle-averaged drift velocity is  
\be{14}
\langle v \rangle
\ = \  \sum_{r=1}^{r_c} f_r \langle v \rangle_r  
\ = \ \left[\sum_{r=1}^{r_c}  f_r \bar{u}_r \right] \ \Omega \delta X_p
\eeq
The normalized occupations satisfy 
$\sum f_r=1$, where ${r=0,1,2,...,r_c}$. 
%
%
For a saturated occupation the $f_r$ are proportional to the phase-space area 
of the filled layers. In the latter case $\langle v \rangle$ 
is merely the average velocity within the occupied 
region. If the whole transporting region is saturated,    
one obtains
\beq
\langle v \rangle \ = \ \frac{1}{2}\left(|v^{(+)}|-|v^{(-)}|\right) \ = \ \dot{X}_{p}
\eeq

It should be realized that the dimensionless amplitude 
of the piston velocity is unity.
Accordingly the condition for the emergence of multiple transporting 
layers is ${\delta u_p > 2}$.
Numerical results for $\langle v \rangle_r$
are presented in \Fig{f3}b, 
and additional plots are available in \Fig{f4}. 
Typically $\langle v \rangle_r$ is dominated by 
the the contribution that comes from the free 
ballistic stage of the motion. This contribution 
can be estimated quite easily (see below).

\section{The drift velocity}
\begin{figure}
\centering

\includegraphics[clip,width=0.74\hsize]{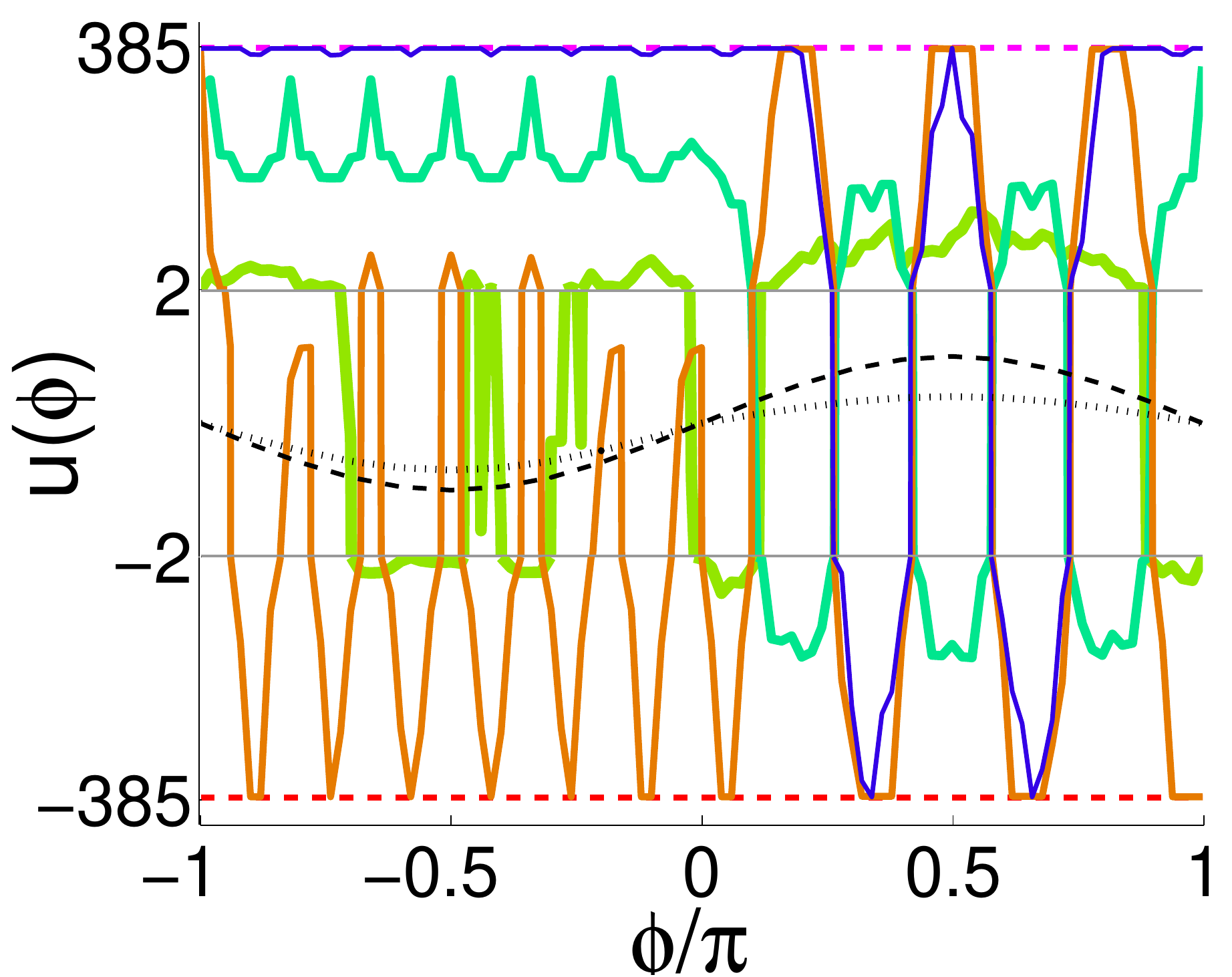} 

\vspace*{2mm}

\includegraphics[clip,width=0.85\hsize]{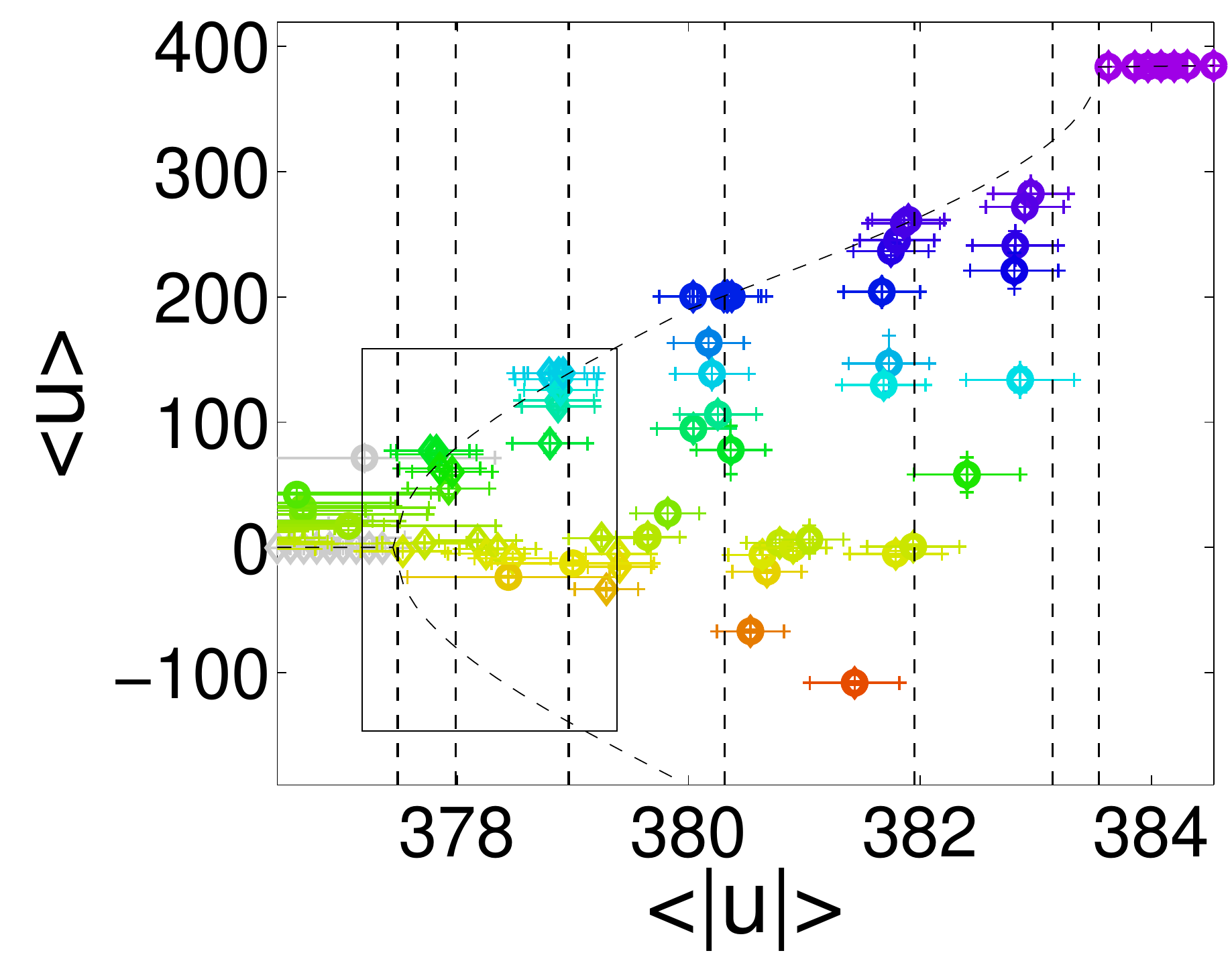}

\caption{
{\bf (Upper panel)}~ 
The drift velocity $u(\varphi)$ for trajectories 
in 4~representative layers (blue, cyan, orange, green) of \Fig{f2}d. 
The motion consists of ``free" and ``adiabatic" stages. 
If the strict adiabatic picture were applicable 
it would be described by an expression $\mathcal{R}(\varphi)\dot{X}_p(\varphi)$, 
with ${0<\mathcal{R}(\varphi)<1}$, as illustrated by the dotted curve.
The dashed curve  is $\dot{X}(\varphi)$.
The dashed horizontal lines correspond to $v^{(\pm)}$.
Note that the ${[-2,+2]}$ range of the vertical axis is zoomed.   
{\bf (Lower panel)}~
The drift velocity $\bra {u} \ket$ 
as function of $\langle |u|\rangle$, 
for the various layers that appear in \Fig{f2}d  (diamonds).
One should exclude the framed data which represent 
additional layers that appear in \Fig{f2}c (circles).   
The dashed black line assumes ${u>0}$ motion during 
the ballistic stage (see appendix), while the vertical dotted lines 
correspond to values of~$u$ that accommodate an even number of bounces.
}

\label{f3}
\end{figure}

\begin{figure}
\centering

\includegraphics[clip,width=0.44\hsize]{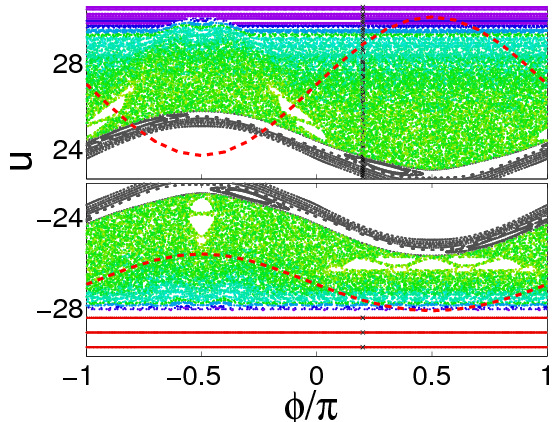}
\includegraphics[clip,width=0.44\hsize]{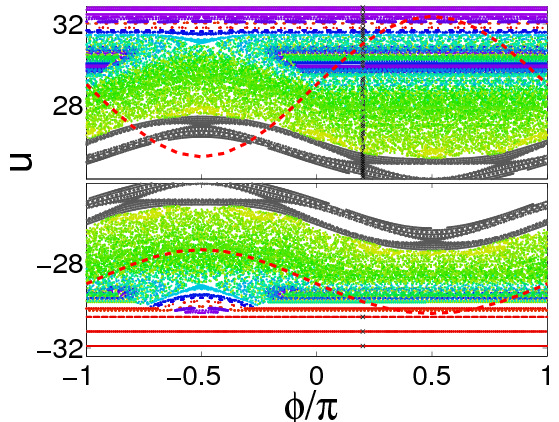}

\includegraphics[clip,width=0.44\hsize]{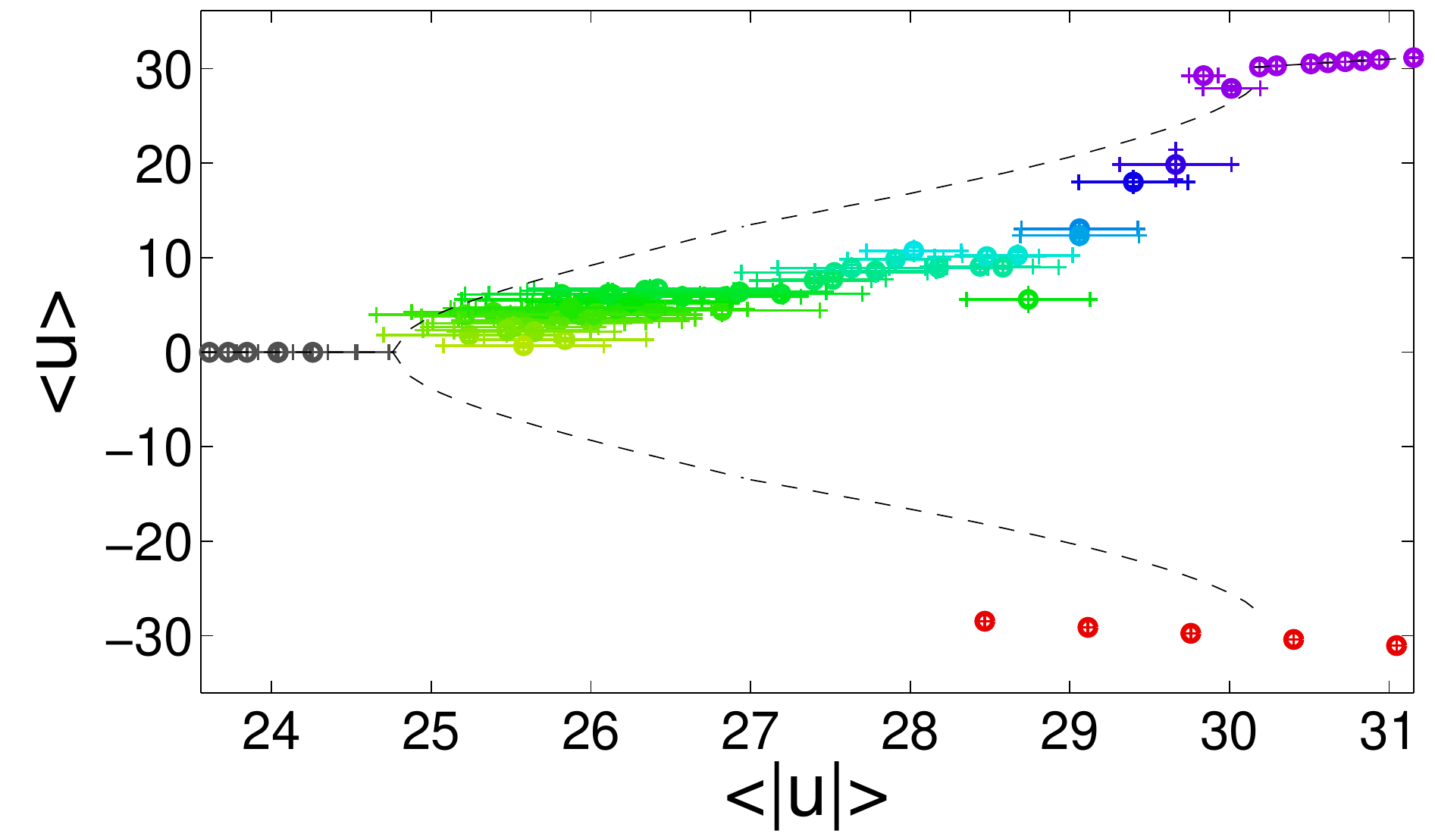}
\includegraphics[clip,width=0.44\hsize]{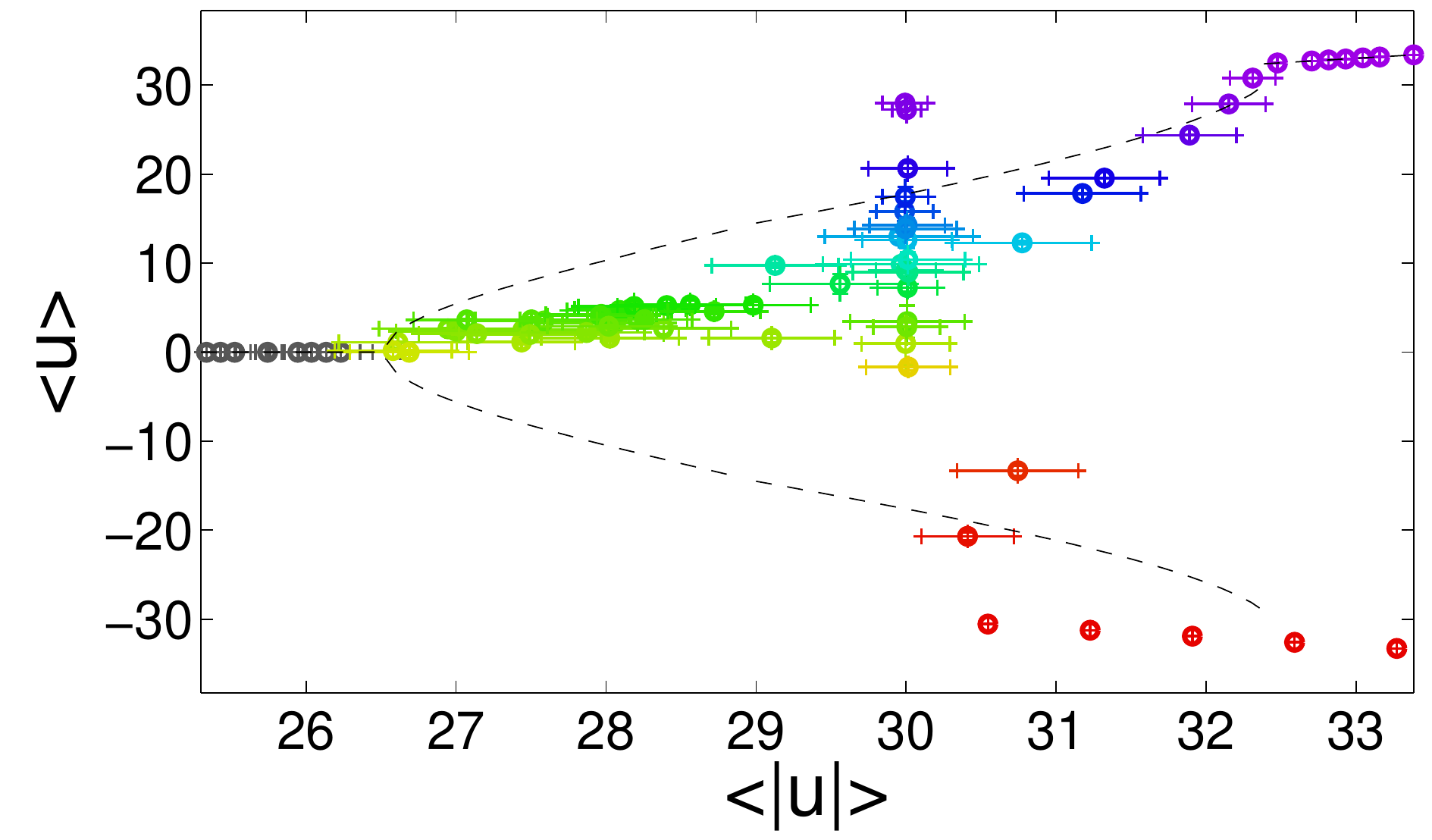}

\caption{
Additional Poincar\'e sections (top) showing mixed phase space. 
There is no barrier. The parameters are $l_{\pm}=15$, 
while  ${u_p=27}$ (left) and ${u_p=29}$ (right)
with ${\delta u_p=2.24, 2.48}$, respectively. 
The color code and lines are as in \Fig{f2}b.
Corresponding plots of $\bra {u} \ket$ in the bottom panels are as in \Fig{f3}.
}

\label{f4}
\end{figure}

At this stage we have to clarify how the drift velocities~$\langle v \rangle_r$ 
are determined by the dynamics. 
The winding number (WN) of a trajectory 
that is generated by Eqs.(\ref{e5}-\ref{e8}), 
and its duration,  are respectively  
\beq
\mbox{WN} = \sum_n g_b(v_n)\mbox{sign}(v_n),  
\ \ \ \ \
\mbox{Time} = \sum_n  \frac{L}{|v_n|} 
\eeq
The drift velocity is 
\be{-1}
\bar{v} &=& \frac{ \mbox{WN} } { \mbox{Time} } L 
= \frac{ \sum_n g_b(v_n)\mbox{sign}(v_n) }{ \sum_n  |v_n|^{-1} } \\
&=&  {\overline{g_b(v_n)\mbox{sign}(v_n)}} \ \Big/ \ {\overline{|v_n|^{-1}}} 
\eeq
Using ergodicity the time average  
can be replaced by phase-space average.
We just have to remember that $\varphi$ 
is like time, which is canonically conjugate 
to the energy. Hence the integration 
measure in  phase-space is $\propto v \df v \df \varphi$. 
Consequently we get the obvious result  
\beq
\bar{v} \ = \ 
\frac{\iint  g_b(v) \, v \, \df v \df \varphi}
{\iint \df v \df\varphi}   
\ \equiv \ \int v(\varphi) \frac{\df\varphi}{2\pi}
\eeq
where the phase-space integration extends 
over the layer that is filled by the trajectory, 
and $v(\varphi)$ is defined as the 
parametric drift velocity.
Results for the drift velocity are presented in \Fig{f3}a.

\section{Estimate of the `free' contribution}
(can be skipped in first reading).
Looking in \Fig{f2}c we observe that in the absence of a barrier 
the motion in the transporting layers is not strictly adiabatic. 
Rather it contains a ``diabatic transition" 
from {\em adiabatic} to {\em free} motion   
when $g_p$ switches from~$0$ to~$1$.  
For ${u>0}$ this transition takes place 
at the intersection of the stochastic layer with ${u^{+}(\varphi)}$. 
The relative filling of the ${u<0}$ branch is determined
by the number of bounces that are accommodated
in the adiabatic stage. It follows that the dependence of
the drift velocity on $\langle |u| \rangle$ is modulated as seen in \Fig{f3}b. 

Neglecting small uncertainty of order $\dot{X}/v$, 
the ballistic motion takes place in the $\Delta\varphi$ interval 
where ${|u|>u_p(\varphi)}$. Hence this interval is determined 
by the roots of the equation ${u_p+\delta u_p\sin(\varphi)=|u|}$.
Assuming that the ballistic motion takes place 
in the $u>0$ region of phase-space, without branching
to the $u<0$ region, we get the upper bound estimate  
\beq
\langle u \rangle_r   &=& \bra \abs{u} \ket \left(\frac{\Delta \varphi }{2\pi}\right)
\eeq
An improved estimate should take the branching 
into account. This branching depends on the 
number of bounces that are accommodated in the 
adiabatic stage. i.e. during the 
interval ${\Delta\varphi_c = 2\pi - \Delta\varphi}$.
The variation of $\varphi$ between successive bounces 
and the number of accommodated bounces are respectively 
\beq
\Delta \varphi_0 = \frac{2 \pi \ell}{\bra \abs{u} \ket}, 
\ \ \ \ \ \ 
m_{\tbox{bounces}} = \frac{\Delta\varphi_c}{\Delta\varphi_0}
\eeq
We expect the branching to become negligible 
each time that $m$ crosses an even integer number.
This in confirmed by the numerical results of \Fig{f3}b.

\section{Discussion of the adiabatic picture}
Canonical adiabatic theory~\mycite{LLbook}
ensures the invariance of the actions for a sufficiently 
slow perturbation of an otherwise integrable system. 
It is analogous to the conservation of the energy-level-index 
in its quantum version~\mycite{Mes}. 
It holds as long as the trajectory does not change its 
topology. If a trajectory does change its topology,  
we call this occurrence a {\em diabatic} transition. 

In the oscillating-piston model, 
each time that the motion switches 
between {\em free} ballistic motion, 
{\em extended} adiabatic motion, 
and {\em bounded} adiabatic motion,    
it is a diabatic transition. 
The two types of adiabatic motions  
are described in the appendix.

The small parameter of the adiabatic theory is~$\dot{X}$.   
In the absence of magnetic field the zero-order adiabatic 
states carry no current. This is true both classically   
and quantum mechanically: note that in the latter case 
the parametric eigen-function ${|X\rangle}$ is real. 
It follows that adiabatic transport requires to go beyond zero-order.
Linear response theory is based on a {\em first-order} treatment,    
leading to the Kubo formula for the transport coefficient. 
In open geometry the scattering formalism leads to the same result.
 
Using a quantum language, but referring on equal footing 
to the classical picture, the evolving zero-order adiabatic 
state ${|X(t)\rangle}$ does not satisfy the continuity equation: 
at any moment ${\langle I  \rangle =0}$. 
Still we can deduce from the zero-order parametric solution 
a non-zero result for the current. 
This is done by associating a parametric velocity 
to each ``piece" of the evolving probability distribution. 
This leads to \Eq{e160}.   
We note that such procedure has been used in \mycite{manz}.
We also note that such procedure becomes ambiguous 
in the case of multiple path geometry: to associate 
a ``displacement velocity" to each piece in phase space 
is not always well defined.

\section{Contrasting with adiabaticity}
The observed results imply that even for very slow driving 
the analysis should go beyond the adiabatic picture. 
If a first-order adiabatic transport picture were applicable,  
the drift velocity of the particle would adjust itself 
to the motion of the piston, and it would be possible  
to write \Eq{e10} with an $\Omega$-independent~${\mathcal{R}}$. 
In practice we observe that $\langle v \rangle_r$ 
has no simple linear relation to $\dot{X}_p$.  
The deterministic adiabatic result \Eq{e160} 
would be obtained if we had a saturated occupation of the 
region bounded by $u_p^{\pm}(\varphi)$. 
It can be regarded as a rough approximation to the total 
adiabatic contribution that we discuss below.
 
We already pointed out that the value of $\langle v \rangle_r$ 
within a layer of phase space is a sum of adiabatic-like  
and free ballistic contributions. The {\em adiabatic}-like contribution 
is proportional to the integral over $\dot{X}_p$ during the ${g_p=0}$ stage. 
The {\em free} ballistic contribution is due to the average velocity 
in the ${g_p=1}$ stage, during which the particle 
circulates the ring without being back-scattered. 
The result of the free ballistic contribution depends on the 
branching that has been explained previously, 
and accordingly, due to the alternating branching ratio, 
we observe an alternating net result for $\langle v \rangle_r$ 
as we go from layer to layer (\Fig{f3}).
  
With a barrier the motion is somewhat more chaotic, 
and the local drift velocity adjusts better to $\dot{X}$, 
as seen in \Fig{f3}a. A {\em strict adiabatic approximation} 
would be applicable if the momentary motion 
(for a frozen piston position) were chaotic, 
with some finite correlation time $\tau_{cl}$.
Then the adiabatic condition would be $\Omega\tau_{cl}\ll 1$.    
This would require to consider a 2D ring, say a Sinai billiard. 
Within this approximation we could define $\mathcal{R}(\varphi)$ such that \Eq{e10} 
would lead to an $\Omega$~free parametric integral for~$Q$.

\section{Contrasting with Boltzmann}
There are two conspicuous differences between our 
results and the expectations on the basis of the Boltzmann picture.
{\bf \ (a)}~In the Boltzmann picture, with ``high" piston, 
in the absence of a barrier, we expect parametric transport 
with $\mathcal{R}=1$, hence obtaining $Q=0$ upon 
integration, implying zero drift velocity.  
{\bf \ (b)}~Including a barrier, the Boltzmann picture 
suggests a non-zero net transport in the same direction 
that is implied by the pumping operation, 
with ${0<\mathcal{R}(\varphi)<1}$. 
This means ${\bar{u}_r>0}$ for all the layers. 
Both~(a) and~(b) are in contradiction to what we observe.  
In particular we point out that the negative $\bar{u}_r$  
characterizing some of the layers is due to the possibility 
of having a free ballistic contribution with branching ratio 
that favors ${u<0}$ motion.

One should appreciate the essential difference 
between the Boltzmann and adiabatic pictures:
Both would agree qualitatively if during the time 
of ${g_p=1}$ the drift velocity were zero. 
Instead it remains constant. 
One realizes that there is an ``order of limits" issue:  
the Boltzmann picture assumes that there is always some infinitesimal 
reflection that allows in the adiabatic limit a randomization of the velocity. 
In contrast to that in the adiabatic picture the reflection 
during the ${g_p=1}$ stage is strictly zero
and hence  ${\langle v \rangle}$ remains constant.

\section{Summary and discussion}
A phase-space based approach for the analysis of stirring 
in a deterministic driven system has been presented. 
Our oscillating-piston model exhibits a layered mixed 
phase-space structure. The determination of the 
drift velocity requires to go beyond a simple parametric 
theory: in general neither an adiabatic nor a Boltzmann 
picture applies. The drift velocity in some layers 
can even have a sign opposite to the current direction 
that would be expected for a strictly adiabatic pumping (``counter stirring''). 
These chaotic layers appear already for slow driving, 
whereas a homogeneously chaotic phase-space, compatible 
with a stochastic picture of stirring, requires a different limit.
It is important to realize that no simple relation 
can be established between a stirring problem 
and its corresponding pumping problem 
(that is, the same driven potential in an open
configuration). Different paradigms are involved.  

A few words are in order regarding the quantum case \mycite{pmt}: 
In the quantum adiabatic limit $\mathcal{R}$
can be calculated  using the Kubo formalism. 
It has a wide distribution, 
which is the ``geometric conductance"  analogue of
universal conductance fluctuations (UCF). 
One can even observe a counter-stirring 
effect ($\mathcal{R}<0$) which would be impossible  
in the strict classical adiabatic picture.
%
%
The manifestation of dynamical localization requires  
significantly longer time scale of coherence, as explained 
in Ref.\mycite{ratchet2} with regard to Hamiltonian ratchets.

\section{Appendix: adiabatic trajectories}

In this appendix we clarify what are the equations 
that describe adiabatic trajectories for the 
pertinent two types of motion of the Fermi-Ulam 
``box model" and its ``ring model" variation.

{\em Bounded adiabatic motion:} 
Consider a particle that is moving back and forth
in a Box of length $L$, between an infinite
barrier and an oscillating piston. 
In the $n$th collision the change in its velocity 
is $-2\dot{X}(t_n)$. Summing over collisions,
approximating by an integral over $\df t/ (2L/v_E)$, 
one obtains ${v(t) =\const - (v_E/L)X(t)}$,    
where the absolute value of the 
velocity $v_E=(2E/\mass)^{1/2}$ is 
assumed to be approximately constant. 
This implies that the equation of the 
adiabatic curve in the Poincar\'e section 
for this type of motion is  
\beq 
u(\varphi) \ = \ \const + \frac{u_E}{2\pi\ell} \cos(\varphi)
\eeq
Note that in our ring setting $\ell$ is $\ell_{+}$ 
or $\ell_{-}$ depending whether it is ${u>0}$ 
or ${u<0}$ trajectory.

{\em Extended adiabatic motion:} 
There is a similar derivation for a particle 
on a clean ring of length $L$. There is no barrier. 
The particle changes direction each time that it 
collides with the moving piston. 
Consequently $v(t)$ changes sign each period,  
and it is convenient to sum the increments pairwise. 
This leads, after an even number of collisions, 
to the obvious result ${v(t) =\const + \dot{X}(t)}$. 
This implies that the equation of the 
adiabatic curve in the Poincar\'e section 
for this type of motion is  
\beq
u(\varphi) \ = \ \const + \sin(\varphi)
\eeq


\section{Acknowledgments} 
This research was supported by the Israel Science Foundation (grant No.29/11), 
and by the Deutsch-Israelische Projektkooperation (DIP).


\clearpage
\end{document}